# Flat bands in long moiré wavelength twisted bilayer WSe$_2$

Zhiming Zhang[1], Yimeng Wang[2], Kenji Watanabe[3], Takashi Taniguchi[3], Keiji Ueno[4], Emanuel Tutuc*[2] & Brian J. LeRoy*[1]

[1]Physics Department, University of Arizona, Tucson, AZ, USA.

[2]Mircroelectronics Research Center, Department of Electrical and Computer Engineering, The University of Texas at Austin, Austin, TX 78758, USA.

[3]National Institute for Materials Science, Namiki 1-1, Tsukuba, Ibaraki 305-0044, Japan.

[4]Department of Chemistry, Graduate School of Science and Engineering, Saitama University, Saitama, 338-8570, Japan

*email: etutuc@mail.utexas.edu, or leroy@email.arizona.edu.

**The crystal structure of a material creates a periodic potential that electrons move through giving rise to the electronic band structure of the material. When two-dimensional materials are stacked, the twist angle between the layers becomes an additional degree freedom for the resulting heterostructure. As this angle changes, the electronic band structure is modified leading to the possibility of flat bands with localized states and enhanced electronic correlations[1–6]. In transition metal dichalcogenides, flat bands have been theoretically predicted to occur for long moiré wavelengths over a range of twist angles around 0 and 60 degrees[4] giving much wider versatility than magic angle twisted bilayer graphene. Here we show the existence of a flat band in the electronic structure of 3° and 57.5° twisted bilayer WSe$_2$ samples using scanning tunneling spectroscopy. Direct spatial mapping of wavefunctions at the flat band energy have shown that the flat bands are localized differently for 3° and 57.5°, in excellent agreement with first-principle density functional theory calculations[4].**

The formation of moiré superlattices (MSL) between two layers of van der Waals materials, with either a twist angle or lattice mismatch between them can dramatically alter the band structure and hence their electronic properties[7–19]. Novel electronic states, such as replica Dirac cones and the Hofstadter butterfly pattern have been observed in graphene/hexagonal boron nitride (hBN) heterostructures[7,16–18]. In a certain range of small twist angles, the folding of the band structure into a mini-Brillouin zone can form flat bands, giving rise to the localization of electronic states and the enhancement of electron-electron interactions[1–6]. When the Coulomb potential surpasses the kinetic energy of the band electrons, strong correlated states such as Mott insulators, or unconventional superconductivity can arise from the flat bands. Recent transport experiments revealed these states for both magic-angle twisted bilayer graphene (MATBG)[13,14].

Apart from graphene-based MSL, twisted bilayer transition metal dichalcogenides (tTMDs) are also predicted to host flat bands at small twist angles for both homobilayers[4,6] and heterobilayers[3,20]. Furthermore, theoretical studies show that owing to the tight confinement of Wannier orbital states in the low energy bands of tTMDs, they can more accurately simulate the Hubbard model[3] compared to the case of MATBG[21,22]. In contrast to MATBG, where the flat bands only form within ±0.2° around a magic angle[23], in tTMDs flat bands are predicted to form over a wide range of angles and their band width monotonically decreases with the twist angle close to 0° and 60° rotation[4], which allows a more versatile platform for the design of flat band devices and wider control of model parameters. Lastly, the spin-valley locking in TMDs leads to a peculiar, electric-field induced transition from trivial to topological insulator between the lowest moiré bands in tTMDs, at small ~2° twist angles[6].

In tTMDs, 0° (AA) rotation has broken inversion symmetry while 60° (AB) rotation has inversion symmetry unlike MATBG which has 60° symmetry such that these twist angles are identical giving an additional degree of control. For heterobilayer TMDs, recent scanning tunneling experiments have shown quantum-confined states related to the sharp peaks around band edges for aligned $MoS_2$-$WSe_2$ heterobilayers[9]. For homobilayer TMDs, optical studies have shown band gap variations due to stacking effects[24] in twisted bilayer $MoS_2$[25–27]. However, a direct measurement of moiré flat bands and localized states in the long moiré wavelength regime (0° < θ ≤ 7° or 52° ≤ θ < 60°) for tTMDs have not been performed yet. In this study we present local characterization of 3° and 57.5° twisted bilayer $WSe_2$ (t$WSe_2$) via scanning tunneling microscopy (STM) and scanning tunneling spectroscopy (STS). We observe spectroscopic signatures of a flat band around the valence band edge for both angles, and directly image the localized nature of this flat band.

An optical image of the 3° device (See Supplementary Information for 57.5° device) and a schematic of our experimental setup are shown in Figs. 1a, b. All measurements were performed in ultra-high vacuum at a temperature of 4.6 K. The samples were fabricated by a dry transfer technique with controlled rotational alignment between the two layers of WSe$_2$[28]. The tWSe$_2$ sits on a bilayer graphene (BLG) flake to provide a conducting substrate for collecting the tunneling current from the STM tip. A hBN flake partially covers the tWSe$_2$ in order to clamp it down and prevent it from rotating from the designed twist angle. Fig. 1c shows the STM topography of the sample, both a graphene-hBN moiré (~11 nm) and the tWSe$_2$ moiré (~6nm) are visible. The graphene-hBN moiré arises from the BLG and the bottom hBN having only a small twist angle between them. In order to extract the precise local twist angle and identify the different stacking configurations of the tWSe$_2$, we filtered out the graphene-hBN moiré with a low-pass filter in Fig. 1d. The high symmetry points are labeled as AA, B$^{W/Se}$, B$^{Se/W}$ and Br, where AA means eclipsed stacking with W atom over W atom, Se atom over Se atom; B$^{W/Se}$ means staggered stacking with W over Se; B$^{Se/W}$ means staggered stacking with Se over W; Br means the bridge that connects neighboring AA sites. Applying an uniaxial heterostrain model[12] and using the closest distance between the AA sites in 3 directions (L$_1$, L$_2$, L$_3$) as input parameters, we find a twist angle $\theta = 3.00°$ and a uniaxial strain $\varepsilon = 0.42\%$ (See Supplementary Information for details about the twist angle and strain determination). This is in excellent agreement with the designed twist angle of 3°. Fig. 1f illustrates the top view and the side view of the different high symmetry stackings that are identified in the tWSe$_2$ MSL.

Figure 2a shows the constant height STS measurements of the local density of states (LDOS) of the 3° device taken on the 4 different high symmetry sites shown in Fig. 1d. For each of these measurements the tip height was stabilized at a bias voltage of -2.4 V and tunnel current of 100 pA. Then the feedback circuit was switched off, a small ac voltage (10 mV) was applied to the bias voltage and the differential conductance dI/dV was measured as a function of bias voltage using lock-in detection. The constant height dI/dV shows a band gap of 2.2 eV for the AA site and 2.1 eV for all the other sites, an energy difference which is mainly contributed by a valence band edge shift of ~80 meV. While the constant height dI/dV can accurately measure the LDOS around the Γ-point in the center of the Brillouin zone, it fails to detect other states with a large parallel momentum, such as the states that near the K-point[29] (See Supplementary Information for details about Brillouin zone location assignments). This can be overcome by employing a constant current spectroscopy method, where the tip height is adjusted by a feedback loop when the effective tunnel barrier changes due to higher parallel momentum. Figure 2b shows the constant current STS measurements of the critical points in the valence band for the four high symmetry sites. For all of these measurements the tunnel current was fixed at 10 pA, and the differential conductance was measured. While the peak at -1.9 V aligned well for all the locations, the features around the band edge shows significant location dependence. The most striking feature are the sharp peaks present at the B$^{W/Se}$, B$^{Se/W}$ and bridge sites, indicating evidence of flat bands in this type of system as predicted by theory[4,6].

To better identify the locations of the critical points at different high symmetry sites, we fit each constant current dI/dV curve with Lorentzian functions (see Fig. 2c for an example at the AA site and Supplementary Information for all fits), and compare their peak positions in Fig 2d. Figure 2e is a schematic band diagram of AA stacked $WSe_2$ that labels the band extrema that are probed with our STS measurements. For each dI/dV curve there are a set of 3 closely spaced peaks between -1.0 V and -1.2 V and another isolated peak near -1.9 V. We first notice that the $\Gamma_v$ point for the AA site is shifted down by ~0.2 V compared with other high symmetry sites, which is consistent with the constant height dI/dV measurements. Other bands only show small shifts between different stackings. There is also a small splitting between the first two Q points in the conduction band and a bigger splitting between the first two K points in the valence band. These splittings are due to the lack of inversion symmetry of the sample, since the in-plane dipole momentum of the top and bottom layers of $WSe_2$ are almost parallel to each other. We note that such a splitting is not present in the inversion symmetric 2H-stacked bilayer $WSe_2$ where the in-plane dipole momentum of the two layers are anti-parallel[30]. The observed spin-orbit-induced splitting (~130meV) in 3° $tWSe_2$ is much smaller than the calculated spin-orbit splitting in a pure AA stacked bilayer $WSe_2$ (~400mV)[4], this is because the formation of a moiré superlattice folds the band structure into a mini-Brillouin zone thus bringing the bands closer together.

The spectroscopy measurements showed sharp peaks at the $B^{W/Se}$ and $B^{Se/W}$ sites and to verify the localization of the flat band wave function at these locations, we measured the LDOS as a function of bias voltage. Figure 3a is the LDOS map at the flat band energy, in excellent agreement with the predicted wave function of the flat band for a relaxed ($\varepsilon \geq 0.3\%$) MSL[4], it features a conductive hexagon enclosing the insulating AA region. On the other hand, LDOS maps at energies away from the flat band (Fig. 3b, c) show the AA regions as triangular bright spots indicating that at these energies the electrons are at the center of the unit cell. The energies in Figs. 3b and 3c correspond to the top of the first and second valence band at the Γ point. The Supplementary Information contains additional LDOS images for all of the band edges identified in the experiment which show similar behavior to the Γ point.

To further explore the formation of flat bands in long moiré wavelength $tWSe_2$, we have also studied a $tWSe_2$ device with a 57.5° twist angle. We note here that unlike the case of twisted bilayer graphene, where rotations by even or odd multiples of 60° yield the same moiré pattern and band structure, a rotation by an odd multiple of 60° leads to a distinct moiré pattern and energy bands in $tWSe_2$. We perform spatially resolved tunneling spectroscopy at constant current along a line (indicated in Fig. 4a) crossing all the high symmetry points. There are a set of states located between -1.1 V and -1.4 V that evolve with position as seen in Fig. 4c. The sharp peak around -1.1 V at the AB site is due to the presence of a flat band. Different from the 3° case, this flat band is isolated from other bands indicating a gap opening below the flat band energy, consistent with theory calculation[4]. There are also multiple sharp peaks between -1.2V and -1.3V, these states are quantum confined states that originate from the triangular potential and lower energy flat bands[31], similar to the aligned hetero-bilayer tTMDs[9]. Figure 4b shows the density of states at the energy of the flat band showing that the state is localized on the AB site unlike the 3° device which has the first flat band localized on the hexagon enclosing the AA region. This difference between the nature of the flat band wavefunctions for 3°, and 57.5° is due to differences in symmetry caused by the stacking as predicted by theory[4].

Our spectroscopic measurements highlight the existence of moiré flat bands in long wavelength $tWSe_2$ originating from the highest valence band at the $\Gamma$ point. In contrast to the MATBG, where the filled flat bands are localized on the AA sites[11,12,15,19], the filled flat band in 3° $tWSe_2$ is localized on the hexagonal network separating the AA sites, while the first flat band in 57.5° $tWSe_2$ are localized on the AB sites. Our results match well with theoretical predications[4], and open up the possibility of probing flat bands in a vast family of small angle tTMDs. Future, gate-tunable experiments could reveal the phenomenology of correlated states in these systems, such as topological insulating states[6], when the flat bands are tuned to be partially filled.

**Acknowledgement:** Work at the University of Arizona was supported by the National Science Foundation under grants DMR-1708406 and EECS-1607911 and the Army Research Office under Grant No. W911NF-18-1-0420. The work at the University of Texas was supported by the National Science Foundation grants EECS-1610008 and DMR-1720595, the Welch Foundation, and Semiconductor Research Corporation. K.W. and T.T. acknowledge support from the Elemental Strategy Initiative conducted by the MEXT, Japan and the CREST(JPMJCR15F3), JST. K. U. acknowledges support from the JSPS KAKENHI Grant No. 25107004.

**Author Contributions:** Z.Z. performed the STM experiments. Y.W. fabricated the samples for STM measurements. K.W. and T.T. provided hexagonal boron nitride crystals, and K. U. provided the $WSe_2$ crystals. E.T. and B.J.L. conceived the experiments. Z.Z. and B.J.L performed data analysis and wrote the manuscript with input from all co-authors.

**Competing Interests statement**: The authors declare no competing interests.

**Main Figure Legends:**

**Figure 1. Stacking configurations of the 3° $tWSe_2$ device. a**, Optical image of the measured device. Blue and gray dashed lines highlight the hBN and BLG flakes, red dashed lines mark the $tWSe_2$ region. **b**, Schematic of the STM setup on the $tWSe_2$ device. **c**, Atomic-resolution STM topography on the 3° $tWSe_2$ sample, probed at fixed bias voltage $V_{bias}$ = -2.5 V and current $I$ = 100 pA. **d**, A zoomed-in view of panel **c** with the graphene-BN moiré filtered out. **f**, Illustration of the different stacking configurations: AA, $B^{W/Se}$, $B^{Se/W}$. The blue and cyan colors denote the W atoms, while pink and yellow denote the Se atoms in the two layers.

**Figure 2. dI/dV spectra on the 4 high symmetry points for the 3° $tWSe_2$. a**, Constant height mode dI/dV vs. bias voltage data, acquired at $I$ =100 pA, probing states in both conduction and valence band. **b**, Constant current mode dI/dV vs. bias voltage, acquired at $I$ =10 pA, probing states in the valence band. **c**, Constant current dI/dV vs. bias voltage data measured at the AA point (black), along with fitting functions (red). **d**, Bias voltage values at fitted peak positions at the four high symmetry points. **e**, Schematic of the band structure of AA stacked $WSe_2$ with all of the measured band extrema labeled. **f**, A zoom-in plot of **d** around the valence band edge.

**Figure 3. Spatially resolved LDOS at different energy. a**, LDOS maps at the flat band energy ($V_{bias}$ = -1.1 V), acquired at $I$ = 10 pA. **b** and **c**, LDOS maps at the Γ energies that are away from the flat band ($V_{bias}$ = -1.4 V and -1.8 V respectively), acquired at $I$ = 10 pA.

**Figure 4. Spatially dependent spectroscopy of the 57° tWSe$_2$ device. a**, Topography image with illustration of the line along where the dI/dV spectra was taken. High symmetry locations are marked as AB, B$^{W/W}$ and B$^{Se/Se}$ where AB corresponds to direct stacking with W over Se and Se over W, B$^{W/W}$ corresponds to staggered stacking with W over W and B$^{Se/Se}$ corresponds to staggered stacking with Se over Se. **b** LDOS maps of the flat band, acquired at $V_{bias}$ = -1.09 V, $I$ = 50 pA. **c**, Line cut showing constant current dI/dV spectra, acquired at $I$ = 50 pA.

## Methods

Sample Fabrication

Our tWSe$_2$ sample was fabricated by sequential pickup steps using a hemispherical handle substrate with rotational control. Starting with a large single grain monolayer WSe$_2$ trimmed into two separate sections by plasma etching, we sequentially picked up the two sections by a BLG-hBN heterostructure attached to the hemispherical handle, with the BLG in direct contact to tWSe$_2$. Between the first, and second WSe$_2$ section pick-up, the substrate was rotated by 3° or 57.5° to create the tWSe$_2$. Another hBN flake partially covering the tWSe$_2$ was subsequently picked up in order to clamp the tWSe$_2$ down and prevent rotation from the designed twist angle. The stacking structure was then placed on a SiO$_2$/Si substrate with tWSe$_2$ and hBN clamp on top. Metal electrodes were defined and deposited with Ni/Au to complete the device.

STM Measurements

STM/STS measurements were performed in the ultrahigh-vacuum LT-STM (Omicron) operating at 4.6 K. dI/dV spectroscopies were acquired by adding a 10 mV a.c. voltage at 617 Hz to the bias voltage and measuring the current with lock-in detection. dI/dV spectroscopy was performed in two different modes, a constant height mode where tip height was stabilized at a particular bias voltage and the feedback circuit was turned off while ramping the bias voltage. A second mode was a constant current mode, where the current feedback was left on while the bias voltage changed allowing the tip to change its height. Electrochemically etched tungsten tips were used for imaging and spectroscopy. All the tips were first checked on an Au surface to ensure that they had the proper work function based on the decay of the tunnel current with distance from the sample. In addition, dI/dV spectroscopy was performed on the Au surface to ensure that the tip had a constant density of states.

**Data Availability**

The data that support the findings of this study are available from the corresponding authors on reasonable request.

Figure 1

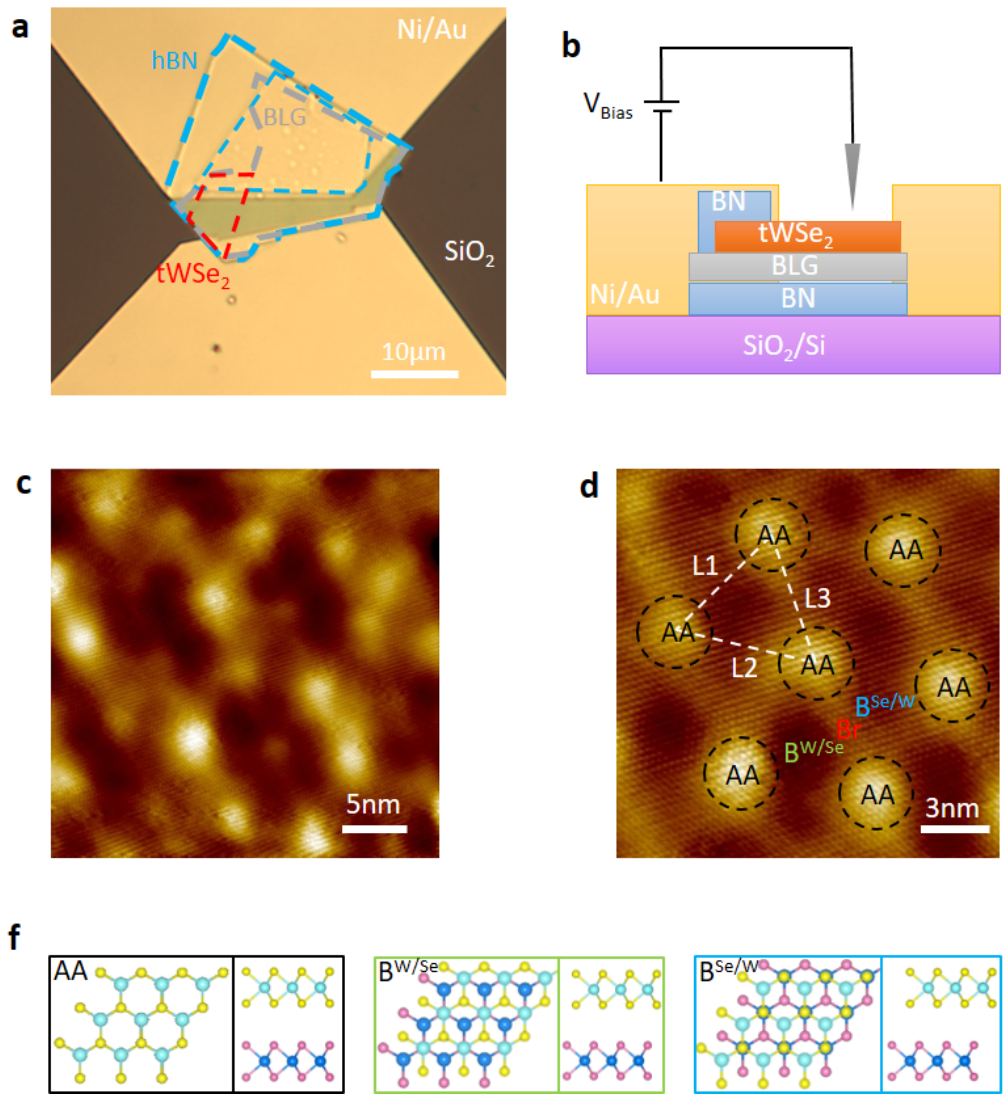

**Figure 2**

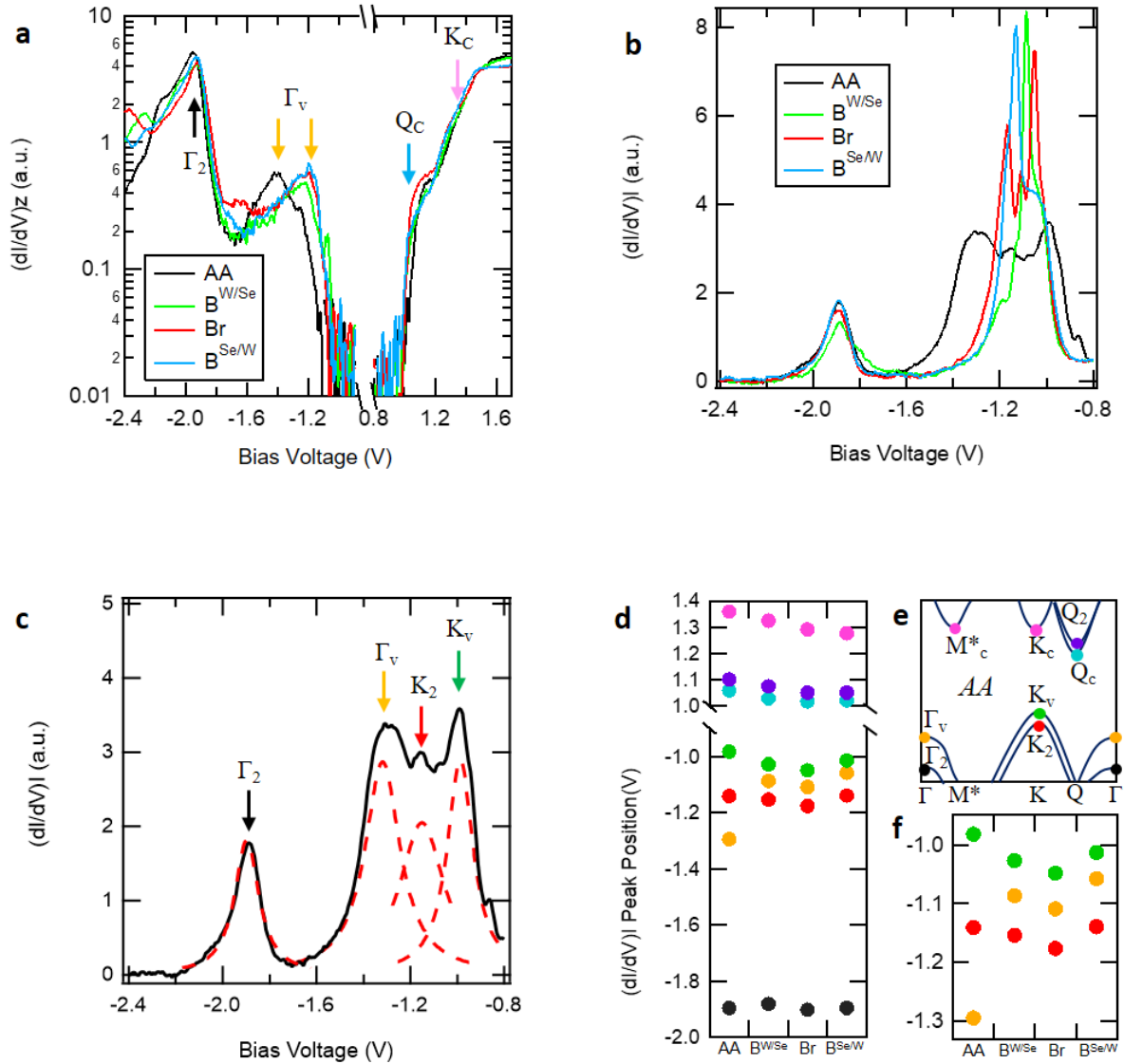

**Figure 3**

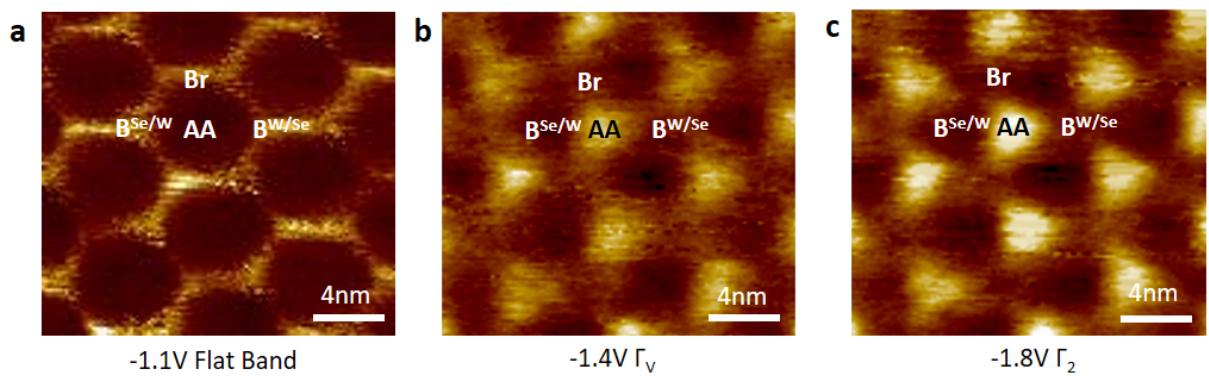

**Figure 4**

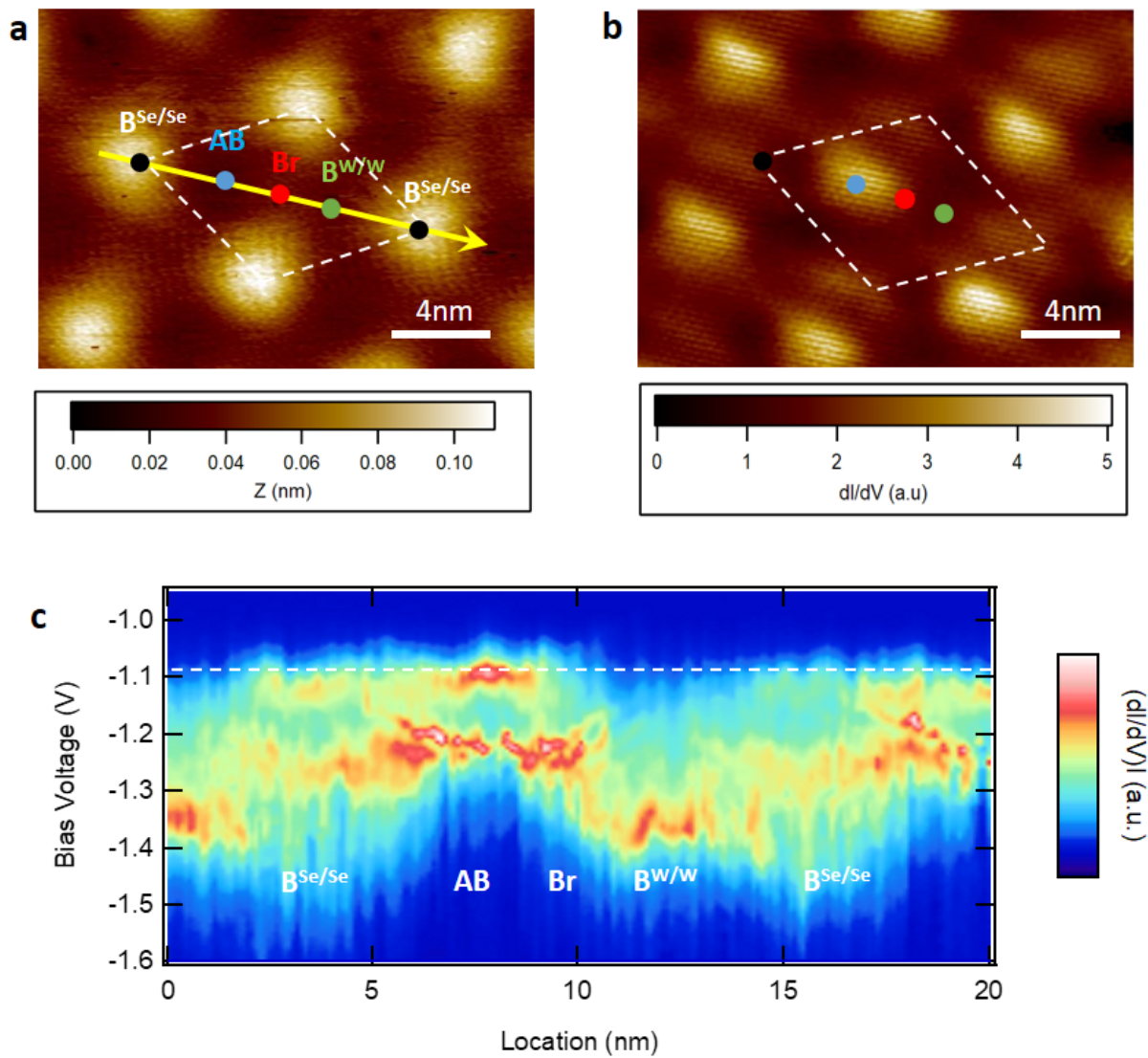